\documentclass[aps,10pt,twocolumn,groupedaddress,superscriptaddress,prl,nofootinbib]{revtex4-2}
\usepackage{amsmath}
\usepackage{mathtools}
\usepackage{amssymb}
\usepackage{graphicx}
\usepackage{textcomp}
\usepackage{bm}
\usepackage{soul}
\usepackage{cleveref}
\usepackage{bbold}

\DeclareMathOperator{\Tr}{Tr}
\DeclareMathOperator{\Real}{Re}
\newcommand{\braket}[1]{\langle{#1}\rangle}

\usepackage[usenames,dvipsnames]{color}

\usepackage[braket, qm]{qcircuit}

\begin{document}

\title{Physical consequences of Lindbladian invariance transformations}

\author{Yohan Vianna de Almeida}\email{yohanvianna@pos.if.ufrj.br}
\affiliation{Instituto de F\'isica, Universidade Federal do Rio de Janeiro, Rio de Janeiro 21941-972, Brazil}

\author{Fernando Nicacio}\email{nicacio@if.ufrj.br}
\affiliation{Instituto de F\'isica, Universidade Federal do Rio de Janeiro, Rio de Janeiro 21941-972, Brazil}

\author{Marcelo F. Santos}\email{mfsantos@if.ufrj.br}
\affiliation{Instituto de F\'isica, Universidade Federal do Rio de Janeiro, Rio de Janeiro 21941-972, Brazil}

\pacs{xxxx, xxxx, xxxx}                                         
\date{\today}
\begin{abstract}
On its own, the invariance properties of Markovian master equations have mostly played a mathematical or computational role in the evaluation of quantum open system dynamics. Because all forms of the equation lead to the same time evolution for the state of the system, the fixation of a particular form has only gained physical meaning when correlated with additional information such as in the evolution of quantum trajectories or the study of decoherence-free subspaces. Here, we show that these symmetry transformations can be exploited, on their own, to optimize practical physical tasks. In particular, we present a general formulation showing how they can be used to change the measurable values of physical quantities regarding the exchange of energy and/or information with the environment. We also analyze examples of optimization in quantum thermodynamics and, finally, discuss practical implementations in terms of quantum trajectories.
\end{abstract}

\maketitle

In physics, invariance properties can generally be explored to discover new traits of a certain system as well as to gather deeper insights on the phenomena behind it. In many cases, they are related to symmetries of the system either derived from fundamental principles, such as in Noether's Theorem~\cite{arnold}, or from the redundancy of degrees of freedom in the fundamental theory, such as the derivation of the electromagnetic fields from a vector and a scalar potentials~\cite{jackson}. 
 
They can also arise from a lack of knowledge over the detailed structure of the system or of its interaction with an external environment. In the second case, whenever the Markovian conditions are met, the dynamics of the system is governed by a master equation in the GKSL form, which is invariant over some symmetries transformations~\cite{BreuerPetr}, here referred to as Lindbladian Invariance Transformations (LITs). 

These transformations, that preserve the average time evolution of the system, are well known and have been exhaustively examined since the first derivations of the GKSL equation. They are particularly relevant when measurements are allowed either on the system or the environment. In fact, the chosen measurements and the corresponding time evolution of single realizations of the dynamics are directly related to each other and to specific forms of the equation. In this case, it has been shown that intrinsic properties of the system such as coherence~\citep{coherence1, coherence2}, entanglement~\citep{entang1,entang2,entang3} and quantum computational power~\cite{comppower1} may be affected by the adopted form. Examples are widely found in the topic of quantum control~\citep{trajectories1,trajectories2,trajectories3}, where proper monitoring strategies are used to drive a quantum state through specifically designed trajectories in time~\cite{trajectories}. 

In general, however, LITs have mostly played a mathematical or computational role in calculating open system dynamics and have only gained physical meaning when correlated with additional information, as mentioned above, or for the discovery of specific set of states such as decoherence-free subspaces. In all those years and to the best of our knowledge, it has rarely been discussed whether LITs could have physical consequences on their own and if those consequences could be exploited in practical tasks. 

A recent work has raised this question by showing that, for systems exchanging energy with its environment, the usual definitions of heat current and injected power are not invariant by LITs~\citep{gaugethermo1}. Here, we go beyond a purely academic discussion to try to understand and exploit the physical consequences of LITs on their own and to study how they can be used in the optimization of specific quantum tasks. We also discuss possible physical implementations of the proposed protocols in terms of quantum trajectories and exemplify our ideas in low-dimensional quantum systems by investigating the optimization of the energy flux into the system as well as its asymptotic stored ergotropy, both physical quantities that are not invariant under LITs.

\subsection{Dynamics invariance and LIT strategies}
We begin by assuming a $d-$dimensional system $\rm S$ subject to a free Hamiltonian $H_0$ and coupled to its environment $\rm E$ in such a way that its dynamics is given by the following Markovian master equation:

\begin{equation}
\dot{\rho}(t) = -i[H,\rho(t)] +  \sum_\mu \left(L_{\mu}\rho(t) L_{\mu}^{\dagger} - \frac{1}{2}\{L_{\mu}^{\dagger}L_{\mu},\rho(t)\}_+\right),
\label{LindbladME}
\end{equation}

\noindent where $\{A,B\}_+$ is the anti-commutator and we make $\hbar = 1$ for simplicity. In this equation, the interaction between $\rm S$ and $\rm E$ is manifested both as a correction of energy in $\rm S$, already comprised in $H$, and a non-unitary part given by the so-called Lindblad operators $L_{\mu}$. Even though the set $\{H,L_{\mu}\}$, which we henceforth call a \textit{strategy}, derives both from the quantum state of $\rm E$ and the specificity of its interaction with $\rm S$, equation (\ref{LindbladME}) only describes the effective action on $\rm S$ due to the presence of the environment. In many physical applications, the details about $\rm E$ are either unknown or irrelevant and this effective description is all we need in order to evaluate the time evolution of meaningful physical quantities in the system.

This lack of knowledge, however, means that the set of operators $\{H,L_{\mu}\}$ is not uniquely fixed, leaving equation (\ref{LindbladME}) unchanged under \textit{Lindbladian Invariance Transformations} (LITs). In fact, one may define new operators

\begin{equation}
L_{\mu} \rightarrow L'_{\mu} = U_{\mu\nu}L_{\nu} + \Gamma_{\mu}, 
\label{Lprime}
\end{equation}

\noindent with implicit summation on $\nu$, which will maintain the evolution of $\rho (t)$ invariant provided $H$ is transformed accordingly: $H \rightarrow H' = H + \delta\!H$, where

\begin{equation}
\delta\!H = \dfrac{1}{2i}\left(\Gamma_{\mu}^{*}U_{\mu\nu}L_{\nu} - \Gamma_{\mu}U_{\mu\nu}^{*}L_{\nu}^{\dagger}\right) + \phi.
\label{Hprime}
\end{equation}

\noindent Here, $U_{\mu\nu}$ are the entries of a unitary matrix, $\Gamma_{\mu}$ are arbitrary complex numbers, $\phi\in\mathbb{R}$ and, in general, all these parameters can be time-dependent. Note that the LIT invariance has an important physical consequence, since it shows that, for any given time evolution of S, there are infinitely many different ways to define how energy and information is exchanged with $\rm E$, encoded in the changes on $L_\mu$, and to split the interaction energy between them, represented by the corresponding transformation in what is considered the system's energy $H$.

It is straightforward to show that the set of transformations $\left\{H, L_{\mu}\right\} \rightarrow \left\{H',L'_{\mu}\right\}$ defines a group action operation which maintains the evolution for $\rho (t)$ unchanged, \textit{i.e.}, exclusive functions of $\rho(t)$ cannot distinguish between the strategies $\left\{H,L_{\mu}\right\}$ and $\left\{H',L'_{\mu}\right\}$. Nonetheless, quantities that also depend on $\{H,L_\mu\}$ may not be invariant under LITs and a natural question arises: can one exploit this non-invariance to probe relevant phenomena or to enhance physical protocols?

For this to happen, these quantities have to rely on specific partitions of the master equation. In other words, consider a physical quantity $\mathcal F(t) = {\mathcal F}[\{ H(t), L_\mu(t)\}, \rho(t), O(t)]$ defined as a functional of the system's operators $\{ H(t), L_\mu(t)\},\ \rho(t)$ and eventually of one or more LIT-invariant operators $O(t)$. If $\mathcal{F}(t)$ is strategy-dependent, it means that different sets $\{H'(t),L_\mu'(t)\}$ may generate ${\mathcal F}'(t)={\mathcal F}[\{ H'(t), L_\mu '(t)\}, \rho(t), O]  \ne {\mathcal F}(t)$ for the same $\rho(t)$. In this case, the value ${\mathcal F}' $(t) will explicitly depend on the strategy parameters $\{U_{\mu\nu}(t), \Gamma_\mu(t), \phi(t)\}$ and a specific set of parameters can be chosen in order to optimize ${\mathcal F}' (t)$ for some task.

Once the physical scenario is established and the mathematical optimization problem solved, the next step is to find how to experimentally implement the optimal strategy. The particularities of the implementation depend on the limitations and specificities of each platform. For instance, if one is able to continuously measure the target system, then, under some approximations, the single realization evolution of the system may be described by a Belavkin equation~\citep{belavkin1,belavkin2}. On the other hand, if the environment is accessible for continuous monitoring, then it is unravelled into a stochastic quantum state diffusion (QSD)~\citep{qsd1,qsd2} or quantum jump trajectories, depending on whether continuous or discrete degrees of freedom of the reservoir are observed. As an example, we proceed to connect the latter to the choice of strategy.

If discrete changes in the environment are continuously monitored, Eq. (\ref{LindbladME}) is unravelled into quantum trajectories by expanding the time variation of $\rho(t)$ into small finite time steps \textit{dt} and rewriting the equation in the form $\rho (t+dt) = J_{0}\rho (t) J_{0}^{\dagger} + J_{\mu}\rho (t) J_{\mu}^{\dagger}$~\cite{unravelling}. The so-called Kraus jump operators $J_{\mu} \equiv \sqrt{dt}L_{\mu}$ identify what happens to the system when a particular change $\mu$ is detected in the environment in the interval $dt$, while $J_{0} = \mathbb{1} - i H dt - \frac{dt}{2}L_{\mu}^{\dagger}L_{\mu}$ is applied when no change is detected. The set $\{J_0,J_\mu\}$ corresponds to a POVM applied to the system in each time step $dt$ and the average over all possible outcomes of the environmental measurements recovers the time evolution of the system's state: $\rho (t + dt) = \sum_{i} \frac{J_{i}\rho (t)J_{i}^{\dagger}}{p_{i}(t)}$, with $p_{i}(t) = \Tr \left[J_{i}^{\dagger}J_{i}\rho (t)\right]$ the probability that the jump $J_{i}$ occurs at time $t$.

There is a direct connection between different monitoring strategies and the LIT invariance of Eq. (\ref{LindbladME}). The most general redefinition on the POVM corresponds to changing the set of jump operators as $J_{\mu} \rightarrow J'_{\mu} = U_{\mu\nu}J_{\nu} + \sqrt{dt}\Gamma_{\mu}$. The transformation of the set $\{J_\mu\}$ is equivalent to that found in Eq. (\ref{Lprime}) and the invariant master equation is obtained when averaging out on the modified POVM as long as the no-jump operator is modified accordingly, $J_{0} = \mathbb{1} - i H' dt - \frac{dt}{2}L_{\mu}^{'\dagger}L'_{\mu}$, with the new Hamiltonian $H'$ as prescribed in (\ref{Hprime}).

Finally, note that finding the optimal strategy can be a very difficult problem given that it involves optimizing over infinite sets of parameters in each time step of the system's evolution. This task, however, may be simplified by imposing restrictions. For example, some sets may be impossible to implement in a specific experimental setup whereas, sometimes, the task itself may require simpler optimizations. We proceed to show two examples that have been showing up quite often in the context of developing quantum technologies and where the optimization can be carried on.

\subsection{Energy flux in small quantum systems}

A clear-cut example of a strategy-dependent quantity is found when we calculate the time variation of the internal energy of a system evolving according to Eq. (\ref{LindbladME}), defined by $\mathcal{P} = \partial_{t}\Tr\left(H\rho\right)$, where $\mathcal{P}$ stands for the energy flux into or out of the system. Given the dependence of $\mathcal{P}$ with $H$ it is clear that different sets $\left\{H,L_{\mu}\right\}$ will result in different values for $\mathcal{P}(t)$.

In particular, one can ask which strategy maximizes or minimizes the energy flux in the system by calculating $ \mathcal{P}(t)$ for all possible alternatives. In this case a change on the set $\left\{H,L_{\mu}\right\} \rightarrow  \left\{H',L'_{\mu}\right\}$ causes a change of flux $\mathcal{P}' = \mathcal{P} + \delta\mathcal{P}_{G}$ where the extra contribution is

\begin{equation}
\delta\mathcal{P}_{G} = i\braket{[H,\delta\!H]} + \Real {\braket{L_{\mu}^{\dagger}[\delta\!H, L_{\mu}]}} + \langle\partial_{t}\delta\!H\rangle.
\end{equation}

One must then optimize $\delta\mathcal{P}_{G}$ over different parameter sets $\{U_{\mu\nu},\Gamma_\mu,\phi\}$. Furthermore, since each set is associated to a particular environmental monitoring scheme and, hence, to a different family of POVMs acting on the system, once the optimal strategy is chosen, the parameters fixation automatically defines the respective optimal environmental observation setting.

For instance, in the case of a qubit coupled to a thermal bath at inverse temperature $\beta$, the typical description of its dynamics is given by the set $\{H=\tfrac{\omega}{2}\hat{\sigma}_{z}$, $L_{\pm} = \sqrt{\gamma_{\pm}}\hat{\sigma}_{\pm}\}$, where $\hat{\sigma}_{-}$ ( $\hat{\sigma}_{+}$) corresponds to the system losing (gaining) a quantum of energy $\omega$ to (from) the bath at rate $\gamma_-$ ($\gamma_+$) and $\frac{\gamma_+}{\gamma_-}=e^{-\beta \hbar\omega}$. A general transformation following Eqs.\ (\ref{Lprime}) and (\ref{Hprime}) results in $\delta\! H = \textrm{Re}(\alpha) \hat \sigma_x - \textrm{Im}(\alpha) \hat \sigma_y$, with $\alpha =  -i\left(\Gamma_{\mu}^{\ast}U_{\mu +}\sqrt{\gamma_{+}}-\Gamma_{\mu}U_{\mu -}^{\ast}\sqrt{\gamma_{-}}\right)$, a complex number defined by the chosen strategy. The eventual time dependence of the parameters is implied. Note that $\delta\!H$ is defined on the plane perpendicular to $H$ in the Bloch Sphere.

For simplicity, let us now restrict ourselves to the set of parameters where $\alpha (t) = |\alpha |e^{i\theta (t)}$, with $|\alpha |$ constant. After some algebra, one obtains an extra contribution to the variation of $\mathcal{P}$ given by

\begin{equation}
\delta\!\mathcal{P}_{G} = \dfrac{\alpha (t)}{2}\left[i(\omega + \dot{\theta})-\dfrac{\gamma_{+}+\gamma_{-}}{2}\right]\rho_{ge}(t) + h.c.,
\label{FinalFormPGauge}
\end{equation}

\noindent where $\rho_{ij}(t)=\bra{i}\rho(t)\ket{j}$ stands for the coherence of the system ($i\neq j)$ in the eigenbasis of $H$. Note that, in this example, the final form (\ref{FinalFormPGauge}) indicates a physical interpretation to each strategy: when the time variation of the internal energy of the qubit is considered, a general transformation on the set $\{H, L_{\mu}\}$ acts as a transferrer, which transforms the evolution of the initial coherences in the system into a flow contribution at each time $t$. 

In particular, if the system has no initial coherence in the eigenbasis of $H$, then $\rho_{eg}(t) = 0$ throughout the evolution and the choice of strategy is irrelevant. Also note that the choice of the best LIT, encoded in $\alpha (t)$, depends on the phase of $\rho_{ge}(0)$. In fact, one can choose the sign of $\delta \mathcal{P}_{G}(t)$ by simply changing the relative phase between $\alpha (t)$ and $\rho_{ge}(0)$.

\begin{figure}[ht]
\includegraphics[width=\columnwidth]{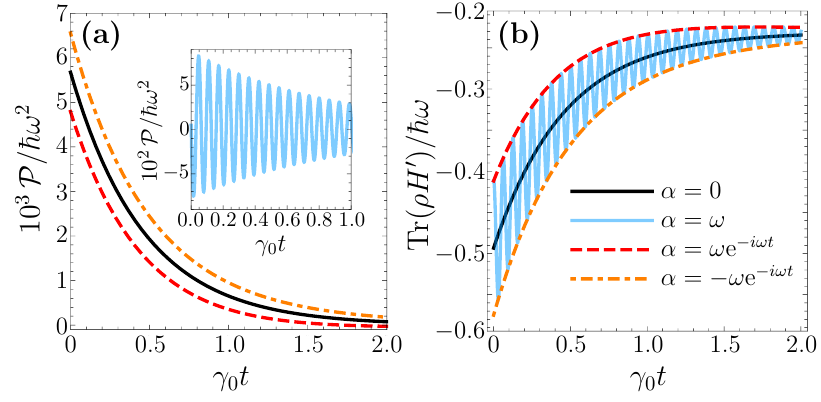}
\caption{
Time evolution of energy flux (a) and internal energy (b) of different strategies leading to the same evolution of a qubit. In all curves, $\gamma_{0} = 10^{-2}\omega$, $k_{B}T_0= 0.2 \hbar\omega$ and $k_{B}T_f= \hbar\omega$.} \label{fig:GaugeTime}
\end{figure}

Finally, Eq. (\ref{FinalFormPGauge}) indicates that one can find monitoring schemes where the exchanged energy per unit of time between the system and its environment can be controlled for any desired task at the expense of any coherence present in the system's initial state.

Figure \ref{fig:GaugeTime} shows the energy flux and internal energy of the qubit for different parameters. The black curve corresponds to $\alpha = 0$, i.e. the original set $\{H,L_\mu\}$. The coloured curves correspond to different sets, where $|\alpha|$ is kept fixed and $\theta$ changes linearly in time. The lightblue curve corresponds to a time independent strategy, where the time evolution of the coherences naturally creates the oscillatory behaviour in the system's internal energy. In this case, $\theta$ only defines the initial phase of the oscillation. On the other hand, by allowing $\theta$ to change in time one is able to coherently interfere either constructively (red, dashed curve) or destructively (orange, dot-dashed curve) with the time oscillations of the coherences of the qubit, accelerating or decelerating the rate at which energy is pumped into the system through the reservoir. In all curves, an initial pure state $\ket{\psi (0)} = \sqrt{p_{g}^{0}}\ket{g} + \sqrt{p_{e}^{0}}\ket{e}$ evolves through the same $\rho(t)$ and asymptotically reaches a Gibbs state $\rho^{\rm as} = p_{g}^{f}|g\rangle\langle g| + p_{e}^{f}|e\rangle\langle e|$ where, $p_{i}^{j}$ is given by a Boltzmann distribution with energy $E_{i} = \langle i|H|i\rangle$ and inverse temperature $\beta_{j}$. We are also considering jump ratios $\gamma_{+} = \gamma_{0}\overline{n}$ and $\gamma_{-} = \gamma_{0}(\overline{n} + 1)$, where $\overline{n} = (1-e^{-\beta_{f}\omega})^{-1}$.

\subsection{Ergotropy on a quantum battery}

Another example of strategy-dependent quantity is the ergotropy stored in a quantum battery~\citep{qbattery1, qbattery2, qbattery3, qbattery4, qbattery5, qbattery6, qbattery7, qbattery8, qbattery9}. Quantum batteries are devices able to store energy in quantum systems and, as the ordinary ones, deliver this energy as useful work~\citep{erg1, erg2, erg3, erg4, erg5}. Ergotropy is the maximum amount of energy that can be extracted from the system via unitary transformations and it is given by: 

\begin{equation}
\mathcal{E}[H,\rho] = \sum_{i}p_{i}^{\rm dec}\left(\bra{p_{i}^{\rm dec}}H\ket{p_{i}^{\rm dec}} - E_{i}\right),
\label{ergotropy}
\end{equation}

\noindent where $p_{i}^{\rm dec}$ are the eigenvalues of $\rho$ labelled in order of decreasing magnitude, while $E_{i}$ are the eigenvalues of $H$ labelled in order of increasing magnitude. Equation (\ref{ergotropy}) makes it explicit that the ergotropy is also a non-invariant physical quantity, as a function of $H$, and therefore it may change as a consequence of a Lindbladian invariance transformation. 

For instance, consider again the same qubit subjected to a thermal bath of inverse temperature $\beta$, for which the asymptotic state, achieved when $t\gg \gamma_\pm^{-1}$, is the same Gibbs state previously described, $\rho^{\rm as}_{ii} = \dfrac{e^{-\beta E_{i}}}{\sum_{j=g,e}e^{-\beta E_{j}}}$. Given that $\rho^{\rm as}_{ee} < \rho^{\rm as}_{gg}$, $\rho^{\rm as}$ has no stored ergotropy in the set $\{H, L_{\mu}\}$, which can be straightforwardly checked from Eq. (\ref{ergotropy}). This is not necessarily true, however, if a different strategy $\{H', L'_{\mu}\}$ is picked to drive the dynamics. The change in the Hamiltonian of the system means that $\rho^{\textrm{as}}$ and $H'=H+\delta\!H$ will not commute anymore and, as a consequence, $\rho^{\textrm{as}}$ may store ergotropy with respect to the new Hamiltonian.

The eigenstates of $H'$ are written in terms of the original eigenstates as

\begin{equation}
\ket{\psi_{+}} = \sqrt{n_1}\ket{g} + e^{-i\theta}\sqrt{n_2}\ket{e},
\end{equation}
\begin{equation}
\ket{\psi_{-}} = \sqrt{n_2}\ket{g} - e^{-i\theta}\sqrt{n_1}\ket{e},
\end{equation}

\noindent with the eigenvalues $E_{\pm} = \pm \dfrac{\omega}{2}[2G(\alpha)+1]$, where $G(\alpha) \equiv \dfrac{1}{2}\sqrt{1+\dfrac{|\alpha |^2}{\omega^{2}}} - \dfrac{1}{2}$, $n_{1} = \frac{G(\alpha)}{2G(\alpha)+1}$ and $n_{2} = \frac{G(\alpha)+1}{2G(\alpha)+1}$. It is very easily checked that for $\alpha \rightarrow 0$ (\textit{i.e.}, no change in the strategy) we recover the original states.

After LIT, the transformed ergotropy is given by

\begin{equation}
\mathcal{E}[H',\rho^{\textrm{as}}] = G(\alpha) \hbar \omega\left[\rho_{gg}^{\textrm{as}}- \rho_{ee}^{\textrm{as}} \right].
\label{newErgotropy}
\end{equation}

Note that the addition of $\delta\!H$ per se does not change the asymptotic value of the internal energy of the system, since the new Hamiltonian $\delta\!H$ is completely off-diagonal in the eigenbasis of $\rho^{\textrm{as}}$, implying that $\textrm{Tr}[\rho^{\textrm{as}} H'] = \textrm{Tr}[\rho^{\textrm{as}} H]$. This indicates that no extra energetic resource is being transmitted into the system by $\delta\!H$. However, $\rho^{\textrm{as}}$ is not a thermal equilibrium state anymore and the energy to keep it out of equilibrium and, therefore, to store ergotropy in it, is injected by the transformed reservoir.

As mentioned before, each set of strategy can be associated to a particular physical implementation of quantum trajectories, each one corresponding to a POVM acting on the system and depending on measurements performed over the environment. In particular, different POVMs for the evolution of qubits in contact with thermal-like reservoirs have been discussed in ~\cite{entang1, comppower1}. There, each jump operator $J_\pm$ can be associated to the detection of a corresponding circularly polarized photon emitted by the system into the environment. Direct photon-detection distinguishing both circularly polarized light modes implement the original set $\{H,L_\mu\}$ on the system. Rotating the polarization of the modes before detecting the emitted photon implements the unitary transformations $U_{\mu\nu}$ over the original $L_\mu$ components of the set, according to equation (\ref{Lprime}). On the other hand, combining each mode with a local oscillator field (a coherent state of amplitude $s_\mu$) at a highly reflective mirror  (of reflectivity $r\sim 1$), implements the addition of the $\Gamma_\mu$ term in the same transformation, where $\Gamma_\mu \propto \sqrt{1-r} \times s_\mu$.

The logic behind it is simple: the rotation of the polarization recombines the modes and therefore the jumps to which they are associated, whereas their combination with the local oscillator at the highly reflective mirror superposes the modes with an external photon source in such a way that a click on a given detector imposes a superposition of a corresponding jump (the photon coming from the system) and identity (the photon coming from the external source) acting on the system. Finally, an external classical field directly driving the $|g\rangle\rightarrow |e\rangle$ transition implements the required $\delta\!H$ to complete the set for the new strategy. 

\section{Conclusion}

In this work, we have proposed a systematic protocol to explore symmetries in quantum Markovian master equations beyond their mathematical elegance. It consists of identifying physical quantities that are not invariant under symmetry transformations of the GKSL master equation and searching for the particular form of the equation that optimizes a given practical physical task. The asymptotic stored energy and the energy flux through the evolution of a low-dimensional quantum system were the two examples depicted here but the protocol is general and any figure of merit that varies according to a chosen form is suitable for optimization. Further examples to be explored in the future include heat exchanged with the environment, entropy production, efficiency of thermal machines, preparation of target oriented asymptotic states, etc.

Our method is a fresh approach into the mathematical symmetries of the Lindblad master equation: in contrast to previous schemes, here the evolution of $\rho (t)$ and, therefore, the information encoded in it through time, is unchanged by design. Rather, we optimize for the ability to better extract a certain physical quantity from a fixed time evolution. From a fundamental point of view, this shows that the same sequence of quantum states can correspond to different trajectories in some parameter space that involves quantities that are not invariant to symmetry transformations. It also means, from a practical point of view, that these transformations can be used to optimize tasks that depend on such quantities. 

This is particularly useful whenever the presence of the environment is unavoidable and one needs to save resources throughout a given fixed dynamics, or when dissipative dynamics are essential to perform a given task. The latter is true, for instance, in protocols that rely on intermediate steps such as optical pumping, electromagnetic induced transparency and coherent population trapping, to name a few. The method here presented is particularly well adapted to these cases. This is certainly a novel approach to an equation that has been extensively studied over the spam of five decades and one that has potential practical aspects. 

The method also allows to identify the resource whose consumption is being optimized. In the examples worked on the paper, energy flux requires initial quantum coherence to be consumed throughout the evolution whereas the storing of ergotropy comes from the athermality of the asymptotic state. The identified resources are consistent with previous results found in the literature~\citep{resources1, resources2, resources3, resources4, resources5, resources6}.

In addition, the unravelling of the GKSL equation in terms of quantum trajectories already points to the experimental way to probe the optimization method. In fact, finding the optimal strategy parameters automatically provides the best set of environmental measurements that must be carried on. This meets direct applications in any area where engineering and control of a given reservoir can be explored to enhance physical protocols, as well as areas where continuous monitoring of the environment is performed in order to feedback control a given evolution. Note that, nowadays, there are plenty of atomic (real and artificial) systems where the conditions for such experiments are met, including atoms coupled to open cavities~\cite{Turchette1,Turchette2,Alexia}, superconducting qubits connected to waveguides~\cite{Tsai}, quantum dots embedded in nanowires~\cite{nano}, NV-centers in photonic crystals~\cite{Hadden}, etc. 

Finally, a first idea on how to generalize this protocol starts by wondering about the effects of letting the LIT parameters assume arbitrary time dependences, which is allowed by the symmetries involved. Furthermore, we also let open the question whether this procedure could be performed on other types of master equations, such as those describing non-Markovian evolutions or even master equations describing classical systems. At last, it may also be of interest to study the relations between these symmetries and their applications with fundamental physical questions related to the thermodynamics of quantum systems.\\

\acknowledgments

\noindent Y.V.A., M.F.S., F.N. are members of the Brazilian National Institute of Science and Technology
for Quantum Information [CNPq INCT-IQ (465469/2014-0)]. Y.V.A. thanks Capes for financial support. MFS is tankful for the financial support of FAPERJ (CNE E-26/200.307/2023) and CNPq 302872/2019-1.

\end{document}